\documentclass[manuscript]{aastex}

\usepackage[amssymb]{SIunits}
\def\fig{Fig.}

\def\Fig{Figure}

\def\sect{Sect.}
\def\sects{Sects.}
\def\Sect{Section}

\def\tab{table}

\def\Tab{Table}

\def\eqn{equation}
\def\eqns{equations}

\def\deg{$^{o}\,$}

\shorttitle{The First VLBI SETI Experiment}
\shortauthors{Rampadarath et al.}

\begin{document}


\title{The First Very Long Baseline Interferometric SETI Experiment}


\author{H. Rampadarath, J. S. Morgan, S. J. Tingay and C. M. Trott\altaffilmark{1}}
\affil{International Centre for Radio Astronomy Research, Curtin University, GPO Box U1987, Perth, WA,
Australia}
\altaffiltext{1}{ARC Centre of Excellence for All Sky Astrophysics (CAASTRO)}
\email{hayden.rampadarath@icrar.org}



\begin{abstract}
The first Search for Extra-Terrestrial Intelligence (SETI) conducted with Very Long Baseline Interferometry (VLBI) is presented. By consideration of the basic principles of interferometry, we show that VLBI is efficient at discriminating between SETI signals and human generated radio frequency interference (RFI). The target for this study was the star Gliese~581, thought to have two planets within its habitable zone. On 2007 June 19, Gliese~581 was observed for 8~hours at 1230-1544~$\mega\hertz$ with the Australian Long Baseline Array. The dataset was searched for signals appearing on all interferometer baselines above five times the noise limit. A total of 222 potential SETI signals were detected and by using automated data analysis techniques, were ruled out as originating from the Gliese~581 system. From our results we place an upper limit of $7~\mega\watt\,\reciprocal\hertz$ on the power output of any isotropic emitter located in the Gliese~581 system, within this frequency range. This study shows that  VLBI is ideal for targeted SETI, including follow-up observations. The techniques presented are equally applicable to next-generation interferometers, such as the long baselines of the \textit{Square Kilometre Array (SKA)}.
\end{abstract}





\section{Introduction}
\label{sec:Intro}

The Search for Extra-Terrestrial Intelligence (SETI) seeks evidence for life in the Universe, through the detection of observable signatures from technologies that are expected to be possessed by advanced civilizations. Such searches are mainly conducted at radio wavelengths \citep{Drake08} and, to a lesser extent, optical wavelengths \citep{RM02}. 

The goal of SETI experiments conducted at radio wavelengths is to detect signals that are intentionally aimed at broadcasting the civilization's existence to others, and/or signals that unintentionally leak from the civilization's communications system \citep{Siemionetal08,Fridman2011}. Intentional signals are expected to be narrow in frequency (1~$\hertz$ to 1~k$\hertz$), while unintentional signals may not be \citep{Tarter01,Tarter04,Siemionetal08,Fridman2011}.

An open question is the frequency range over which to search for extra-terrestrial intelligence \citep{Tarter04,Shostak09,Fridman2011}. The frequency range from 1 to 10~G$\hertz$ is generally preferred, however after decades this range is still very much unexplored \citep{Tarter04}. For leakage signals, the lower end of this range is preferred \citep{Tarter01,Tarter04}. For deliberate signals from extra-terrestrial intelligence, the most common suggestions are the 1420~MHz hydrogen line and multiples of 1420~MHz, including $\pi\,\times\,$1420~MHz (4462.336275 MHz) \citep{Blarietal92,Harpetal10}.

SETI experiments are divided into two categories: (1) sky surveys and (2) targeted searches \citep{Shostak09}. The sky surveys make no assumptions about the location of the SETI signals and perform raster scans of the observable sky. 
Current sky surveys are SETI@home, ASTROPULSE and SERENDIP V conducted with the 300~$\meter$ Arecibo telescope in Puerto Rico \citep{Werthimeretao01,Siemionetal08}. 
Targeted searches sequentially examine specifically chosen stars that are deemed to have a chance of harbouring a planet that could sustain life \citep{Tarter04,Fridman2011}. A list of 17,129  stellar systems that are potentially habitable to complex life forms was compiled by \citet{TT03}.

The currently active \textit{Kepler Space Mission} aims to search for Earth-sized planets in and near the habitable zone of Sun-like stars \citep{Boruckietal10}. As of February 2012, \textit{Kepler} has found over 2,300 planetary candidates, with $\sim$46 candidates within habitable zones \citep{Kepler3}. As \textit{Kepler} begins to confirm these planets and proceeds to discover more Earth-like planets in habitable zones, these planets will be investigated for extra-terrestrial intelligence. 

This paper explores the suitability of using Very Long Baseline Interferometry (VLBI) for targeted SETI from potentially habitable planets. SETI is listed under the \textit{The Cradle of Life} Key Science Project for the \textit{Square Kilometre Array} (SKA, \citet{CR04,Tarter04,Schilizzietal10}). This paper also aims to set a foundation for future VLBI SETI projects, including the use of the long baselines of the SKA. 

In this paper we present the first SETI experiment conducted with VLBI. \Sect~\ref{sec:Target} introduces the target for this study, Gliese~581. \Sect~\ref{sec:VLBI} discusses the advantages of using VLBI for targeted SETI and describes the technique of using VLBI for SETI. \Sect~\ref{sec:The Experiment} describes the observations and data analysis, and presents the results of the experiment. Discussions and conclusions are presented in \sects~\ref{sec:Discussion} and 6, respectively.    

\section{Target}
\label{sec:Target}

The target for this pilot study is the M-dwarf star Gliese~581 (Gl581), located 20 light-years (ly) distant in the constellation Libra \citep{Udryetal07,Vogtetal10}. 

In 2007, \citet{Udryetal07} announced the discovery of a planet on the edge of Gl581's habitable zone. The planet, Gliese~581d (Gl581d), is the fourth planet in the Gliese~581 system. It has a mass of $5\,M_{E}$, an orbital period of 83 days and an orbital semi-major axis of 0.25~AU ($\sim$40~milliarcseconds (mas))\footnote{this is the projected angular distance as seen from Earth}. \citet{Wordsworthetal10} discussed the effects of surface gravity, surface albedo, cloud coverage and CO$_{2}$ density on the surface temperature of Gl581d. Their results showed that Gl581d may have the necessary conditions to sustain liquid water on its surface. Thus, Gl581d may be the first confirmed exoplanet with the possibility to sustain life \citep{Wordsworthetal10}.

\citet{Vogtetal10} claimed to have discovered a planet within the habitable zone of Gl581. The suspected planet, Gliese~581g (Gl581g), is located between Gl581d and Gliese~581c (the third planet). Gl581g has a proposed mass $\sim\,3\,M_{E}$, an orbital period of 37 days and an orbital semi-major axis of 0.15~AU (projected angular distance $\sim$23~mas). The existence of a low mass planet in the habitable zone of Gl581 has been questioned by \citet{Tuomi11}. 

No radio emission was detected from the position of Gl581 by the VLA FIRST survey \citep{FIRST} above a limiting flux density of 1~mJy at 20~$\centi\metre$.


\section{Using VLBI for Targeted SETI}
\label{sec:VLBI}

\subsection{VLBI}
Very Long Baseline Interferometry (VLBI) allows, by the combination (via a correlator) of signals from multiple radio telescopes, the emulation of a telescope the size of the maximum telescope separation, which is generally hundreds to thousands of kilometres. 
The outputs of the correlator are 4-dimensional (baseline, time, frequency, polarisation) datasets consisting of multiple spectral channels and time averages over $\sim$1~second. The primary characteristic of VLBI is its very high angular resolution, which falls in the milli-arcsecond regime, the highest resolution in astronomy. This high resolution makes VLBI ideally suited for high precision astrometry \citep[see][chap.~12]{TMS}, and observation of high energetic compact objects. 

Existing VLBI instruments are able to monitor a wide range of frequencies, from 10~$\mega\hertz$ (the Low Frequency Array, LOFAR: \citet{Stappers11}) to 230~$\giga\hertz$ (The Event Horizon Telescope: \citet{Brodericketal2011}). In addition, current VLBI systems such as the LBA and the European VLBI network (EVN) include some of the most sensitive radio telescopes in the world.

To date there has been little or no use of VLBI in SETI investigations. The SETI project SETI-Italia \citep{Montebugnolietal01} uses data commensally from the 32~m VLBI Medicina telescope. SETI-Korea \citep{Rheeetal10} has proposed the use of existing and new telescopes from the Korean VLBI Network (KVN) for SETI. In addition, \citet{Slysh91} proposed to use VLBI or space-VLBI satellites to measure the angular size of suspected SETI radio sources. Pseudo-interferometric observations were conducted by the SETI Institute, using both the Arecibo Telescope and the Lovell Telescope, Jodrell Bank Observatory between 1200 and 1750~$\mega\hertz$ \citep{Backus02}. The survey searched for signals from nearby stars with appropriate frequency drift and offset caused by the rotation of the Earth.   

\subsection{Radio Frequency Interference (RFI) Mitigation}
\label{sec:RFI}
Radio frequency interference (RFI) is human-made, unwanted radio emission within the telescope's field of view. Potential sources include mobile phones, radars, television signals, FM-radio, satellites etc. Separating RFI from a real SETI signal is one of the most challenging problems faced by SETI projects. Thousands of hours of computing time are spent identifying and removing RFI from the datasets, before one can search for SETI signals. Since its inception, the SETI@home project has detected over 4.2 billion potential signals. While essentially all these potential signals are RFI, it is possible that therein lies a true extraterrestrial signal \citep{Korpela10}. VLBI offers several ways of discriminating RFI signals from SETI signals not applicable to single dish or even short baseline radio interferometers. 

First, VLBI uses \textit{N} telescopes separated by hundreds to thousands of kilometres giving $N(N-1)/2$ baselines, once the data from independent pairs of telescopes are correlated together. This aspect of VLBI is important, as RFI which is not present at both telescopes involved in a baseline does not correlate. An astronomical or SETI compact radio source  within the interferometer field of view will be detected on all baselines. This is also true for RFI if it is detectable by both telescopes in a baseline. However, even correlated RFI can be easily identified with VLBI, as it produces distinct phase variations in both frequency and time in the correlated output as discussed below.

\subsection{VLBI Phase Variations}
\label{sec:PhaseInfo}

The correlated signals from an interferometer are complex quantities (known as visibilities), with amplitude and phase as the fundamental observables. The amplitudes are proportional to the correlated power of the source, while the phase, $\phi$ represents the residual time delay that results from the wavefront of the radio emission arriving at one antenna before the other. At the centre of the field of view (also known as the phase centre), the delay (and hence the phase) is set to zero, via application of a \textit{a priori} time delay, $\tau$, at the correlator. This time delay accounts for many different effects, but primarily simple geometry. The phase, $\phi$ of a radio point source offset from the phase centre can then be related to its position relative from the phase centre, and is given by \eqn~1 from \citet{Morganetal11},

\begin{equation}
\phi = \frac{2\pi\nu}{c}(lu + mv),
\label{eq:deltaphi}
\end{equation}

where $\nu$ is the observing frequency, \textit{l} and \textit{m} are the coordinates of the source in the plane of the sky, relative to the phase centre, and \textit{u} and \textit{v} are coordinates which describe the instantaneous geometry of the projected baseline, with respect to the phase centre. The reader is referred to \citet{TMS} and \citet{SIRA} for details. 
For a finite observational bandwidth, and for the duration of the observation, the radio source in \eqn~\ref{eq:deltaphi} will display a phase change with frequency and time given by,

\begin{equation}
\frac{d\phi}{d\nu} = \frac{2\pi}{c}(lu + mv),
\label{eq:deltaphasefreq}
\end{equation}

\begin{equation}
\frac{d\phi}{dt}= \frac{2\pi\nu}{c}(l\frac{du}{dt} + m\frac{dv}{dt}).
\label{eq:deltaphasetime}
\end{equation}

By measuring the phase change with time and frequency for a given source, its location with respect to the phase centre can be determined using the above equations. Additionally, if the source's position is known, maximum values for both $d\phi/d\nu$ and $d\phi/dt$ can be determined, using maximum values of $u$ or $v$ and $du/dt$ or $dv/dt$.


\subsubsection{Determining the Phase Variations with Fourier Transforms}
\label{sec:FT}

An estimate of $d\phi/dt$ or $d\phi/d\nu$ for a radio source can be determined by the discrete Fourier transform (DFT) of the complex visibilities.
For narrow-band (persistent or intermittent in time) and broad-band signals, DFTs can be applied to the complex visibilities as a function of time to determine $d\phi/dt$ and frequency to determine $d\phi/d\nu$, respectively.

\begin{figure}[h]
\centering
\includegraphics[scale=0.75]{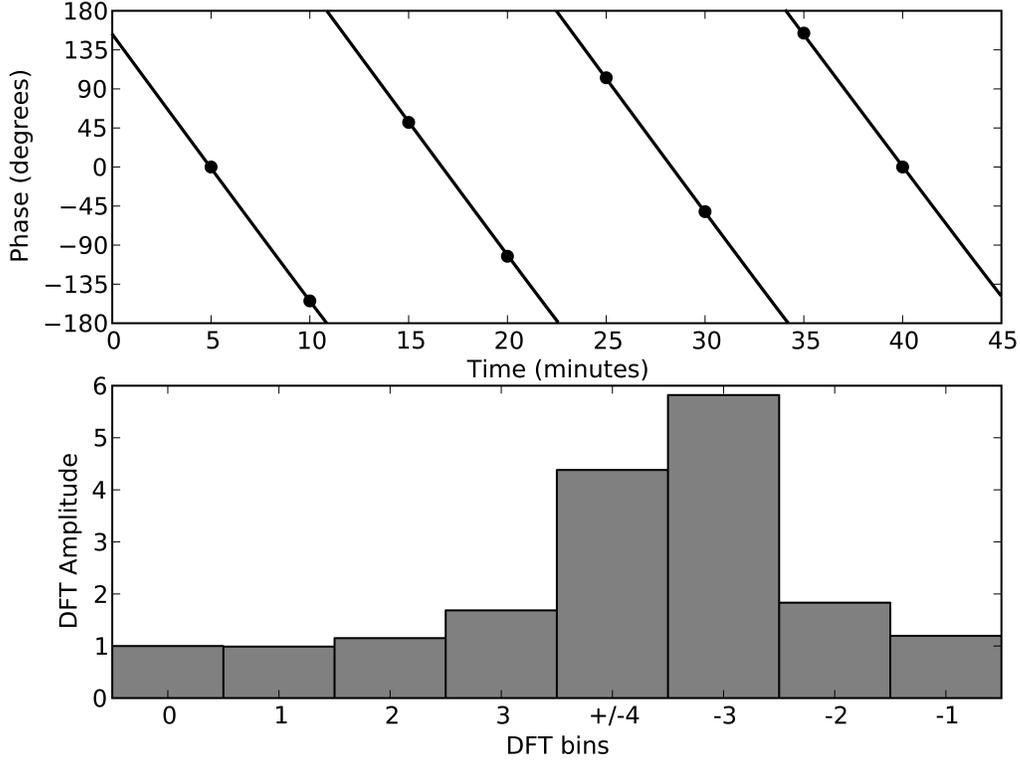}

\caption{The phase variation with time (upper panel), and the corresponding discrete Fourier transform (DFT) for a narrow-band test signal. The DFT bins provide an estimate of the $d\phi/dt$ of the signal. The Nyquist bin at bin $\pm$4, represents the maximum possible $d\phi/dt$, 180\deg per data point. Each data point has a sampling time of 5~minutes. The test signal is found to have a $d\phi/dt$ of $-$135\deg per point, equivalent to 3 turns over the time range. However, due to ambiguities $d\phi/dt$ of this signal can be generalised to $-135\pm180N$\deg per data point, where $N$ is a wide range of positive and negative integers. The same principle applies in determining $d\phi/d\nu$ for broad-band signals, where the DFT is applied to the complex visibilities as a function of frequency.}
\label{fig:FFT}
\end{figure}

This method is illustrated in \fig~\ref{fig:FFT}. The upper panel shows the phase of a narrow-band test signal as a function of time, with a sample time of 5~minutes per point over a 40~minute observation. The lower panel displays the corresponding discrete power spectrum obtained by applying DFTs to the complex visibilities. 

A signal with little or no phase change in time or frequency (implying a source close to the phase centre) will appear in bin zero. However, a signal with high phase change with time (as the test signal in \fig~\ref{fig:FFT}) is indicative of a source far from the phase centre, probably RFI, and will appear in a different bin. The height of the DFT amplitude of bin zero compared with the others gives an indication of the signal to noise of any signal which comes from the phase centre. 

There are, however, ambiguities in signals with $d\phi/dt$ greater than the Nyquist frequency. The component of the test signal located in the $-$3 bin can be generalised as $d\phi/dt$ = $-135\pm180N$\deg per data point, where $N$ is a a wide range of positive and negative integers. The $d\phi/dt$ of the Nyquist bin, due to ambiguities is $\pm$180\deg per data point\footnote{The reader is refereed to \citet{Bracewell} for further details on discrete Fourier transforms}.
 

\section{Searching for Signatures of Extra-Terrestrial Intelligence from Gliese~581}
 \label{sec:The Experiment}
 
\subsection{Observations}
 \label{sec:Obs}
 
\begin{figure}[h]
\centering
\includegraphics[trim=0cm 0.75cm 0cm 1.7cm, clip,scale=0.75,angle=0]{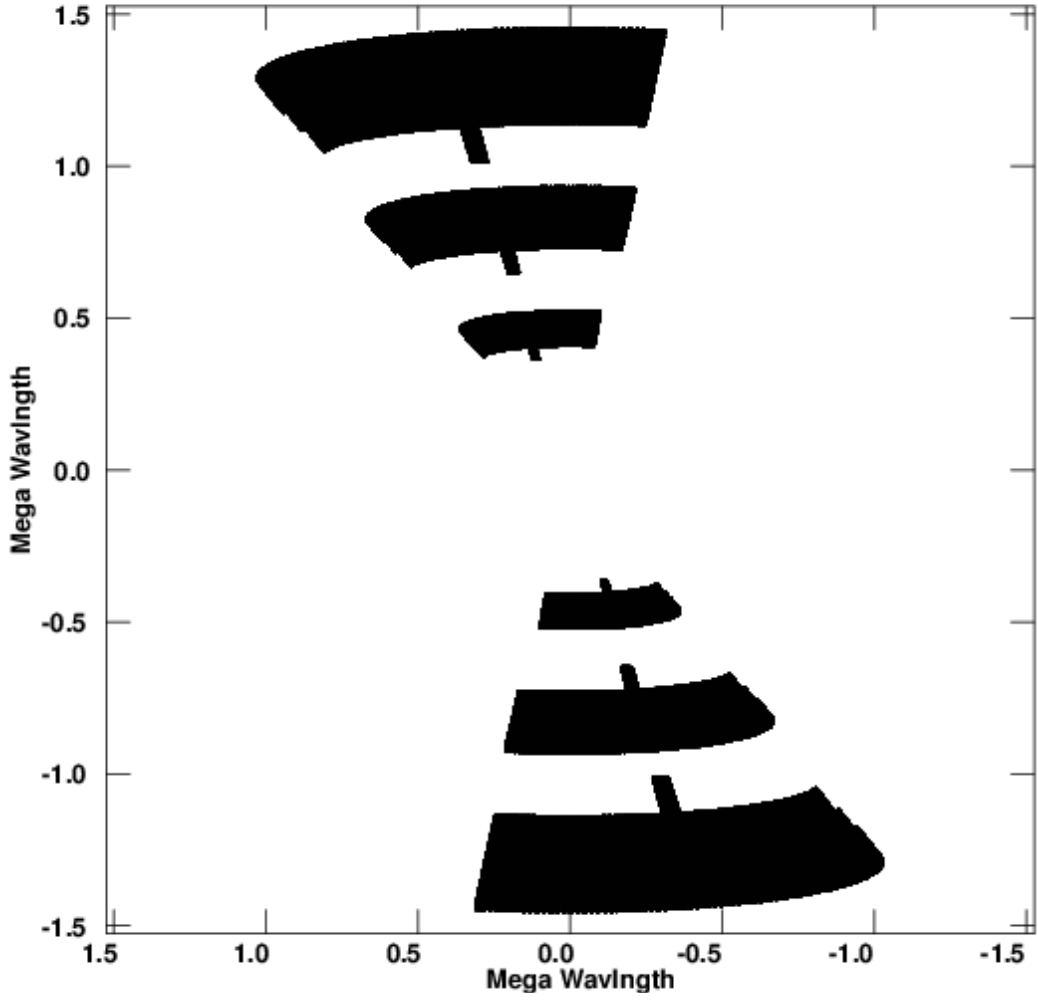}

\caption{uv coverage of the LBA observation of the Gl581 system, in units of mega wavelengths. The x-axis plots the u coordinates, while the v coordinates is on the y-axis. }
\label{fig:uvplot}
\end{figure}

Gl581 was observed on 2007 June 19, for 8 hours in the 20~cm wavelength band by three stations of the Long Baseline Array (LBA): Mopra (Mp), Parkes (Pa) and the Australia Telescope Compact Array (At) in phased array mode. The observation was conducted in three frequency pairs of 64 M$\hertz$ bandwidth centred on: 1262,1312; 1362,1412; 1462,1512 M$\hertz$, in dual-polarisation mode. The uv coverage is shown in \Fig~\ref{fig:uvplot}. This configuration gives a maximum resolution of 127~
mas.
The observations of Gl581 were phase referenced to the ICRF2 source 1510$-$089 (J1512-0905, RA = $15^{h}12^{m}50.532925^{s}$; Dec = $-09^{\circ}05^{\prime}59.82961^{\prime\prime}$, J2000), a calibrator located 2.14 degrees from Gl581 \citep{Feyetal04}. 1510-089, shows an unresolved core with extended structure to the North-East of the core at 2.3~$\giga\hertz$ \citep{FC97}, and 327~$\mega\hertz$ \citep{Rampadarathetal09} on the longest baselines of the Very Long Baseline Array (VLBA). The longest baseline of our observation (At-Pa) is a factor of 9 less than the longest VLBA baseline, hence, 1510-089 is expected to be unresolved in our observations. 

A phase-referencing cycle of 5 minutes (2 minutes on calibrator, 3 minutes on Gl581) was employed. A single observation block consisted of 8$\times$5 minute phase-referencing cycles. The entire observation comprised 12$\times$40~minute observational blocks switching between the three frequency pairs (\tab~1).

The data were correlated using the DiFX software correlator \citep{Delleretal07,Delleretal11}, with a frequency resolution of 1.95 k$\hertz$ per spectral channel, and converted to the standard FITS-IDI format. The lowest and highest 5~M$\hertz$ (2500 channels) were discarded from each band due to band edge drop-off in sensitivity. The data files were then sub-divided into five datasets of 4000 channels each and one dataset of 1268 channels at each centre frequency.


\subsection{Gl581 Coordinates}
\label{sec:Offset}

Due to its proximity to Earth, the exact coordinates of Gl581 at the time of observation must be calculated taking into account parallax and proper motion.
An accurate position was computed via a python script (James Miller-Jones, ICRAR, private communication), using a reference position of Gl581 (RA and Dec) and the barycentric coordinates of the Earth (in AU) at the time of the observation (obtained from NASA's Jet Propulsion Laboratory HORIZONS on-line solar system data and ephemeris computation service\footnote{http://ssd.jpl.nasa.gov/horizons.cgi}).

The reference position for Gl581 was taken from the \textit{Hipparcos} catalog \citep{Perrymanetal97}. Gl581 was observed on 1991 March 1, by the \textit{Hipparcos} satellite, and a position of RA(J2000)=$15^{h}19^{m}26.825^{s}$; DEC(J2000)=$-07^{\circ}43^{\prime}20.210^{\prime\prime}$, was recorded with positional errors: $e_{RA}$=1.99~mas and $e_{DEC}$=1.34~mas. Additionally, a parallax of 159.52 $\pm$ 2.27~mas was measured by \textit{Hipparcos}. A proper motion of  RA = $-$1227.67 $\pm$ 3.54~mas/yr; DEC = $-$97.78 $\pm$ 2.43~mas/yr, was also taken from the \textit{Hipparcos} catalog. The position calculated for Gl581 at the VLBI observation date, is RA(J2000)=$15^{h}19^{m}26.206^{s}$; DEC(J2000)=$-07^{\circ}43^{\prime}17.555^{\prime\prime}$. This calculation has a positional uncertainty of 53~mas, largely due to the cumulative effect of the uncertainty in the proper motion measured by the \textit{Hipparcos} 1991 observation. 

\subsection{Data Reduction and Calibration}

The data reduction and calibration was performed using the National Radio Astronomical Observatory (NRAO) software package \textit{AIPS}\footnote{The Astronomical Image Processing System (AIPS) was developed and is
maintained by the National Radio Astronomy Observatory, which is operated by
Associated Universities, Inc., under cooperative agreement with the National
Science Foundation.
}. Prior to calibration and data analysis, the phase centre of the calibrator was shifted, via the \textit{AIPS} task \textit{UVFIX}, to match the calculated positional difference. Flagging and phase-only calibration (fringe-fitting and self-calibration) were carried out on the phase centre shifted calibrator. The phase solutions were transferred and applied to the Gl581 datasets. After calibration, further analysis was carried out using the \textit{ParselTongue} scripting language \citep{Kettenis2006} and the numerical python package, \textit{numpy}\footnote{http://numpy.scipy.org/}. The flagging, phase calibrations and transfer of solutions were implemented separately for the frequency pairs listed in \Tab~1. 

\subsection{Expected Upper limits to $d\phi/d\nu$ and $d\phi/dt$ for the Gl581 system}

\label{sec:phase_estimates}
The uncertainty in the position of Gl581 (53~mas) and the semi-major axes of the planets (Gl581d = 40~mas and Gl581g = 23~mas) produces a quadratic error of 66~mas and 57~mas on the positions of Gl581d and Gl581g, respectively. The positional uncertainty places upper limits on the $d\phi/d\nu$ and $d\phi/dt$ expected for a SETI signal from Gl581. From \eqns~\ref{eq:deltaphasefreq} $\&$ \ref{eq:deltaphasetime} the upper limits are: $(d\phi/d\nu)_{max}\,=\,(1.2\,\times\,10^{-10}$)\deg per spectral channel\footnote{1 spectral channel = 1.95~$\kilo\hertz$} and $(d\phi/dt)_{max} = (2\,\times\,10^{-4}$)\deg per second (or ($6\,\times\,10^{-2}$)\deg per data point, after averaging each scan).

\subsection{Analysis}
\label{sec:Analysis}
\subsubsection{Search for Candidate SETI Signals}
\label{sec:Search}

After correlation and calibration our dataset consisted of 8 hours of data for the three baselines, divided into 12$\times$40 minute observational blocks. Each 40~minute observation block was divided into three~minute scans of Gl581, consisting of two second time-averaged complex visibilities and 32768~spectral channels. Before any further analysis was performed, each scan was vector-averaged along the time axis (for three minutes) leaving the observational blocks as arrays of data, with 32768 channels and 8 time integrations. 

Since the star Gl581 is not a radio source (see \sect~\ref{sec:Target}), the dataset was not expected to contain any continuum emission. The dataset therefore consisted almost entirely of Gaussian noise at the thermal noise level with a few narrow spectral lines, possibly transient RFI. Any SETI signals present are expected to be weak compared to RFI. For data containing only weak signals (signal-to-nose ratio, SNR $\ll$ 1), the mean of the amplitudes of the complex visibilities can be used as a proxy for the thermal noise \citep[Sect.~9.3]{TMS}. However, the mean is biased by RFI. To avoid this bias, the median of the amplitudes
of the complex visibilities was used in place of the mean as a proxy for the thermal noise.

\subsubsection*{Broad-band signals}

Each vector-averaged scan was searched along the channel axis for visibilities with sufficient amplitudes to have a signal-to-noise ratio (SNR) $>$~5. Detected signals were then checked to see if they occur on all baselines. Signals found on all baselines were plotted as a function of channel and searched for instances of five or more adjacent channels with amplitude and phases displaying characteristics of broad-band signals. There were 22 groups of signals satisfying these requirements, selected as broad-band SETI candidates. 

\subsubsection*{Narrow-band signals}

The observational blocks, consisting of 32768 channels and 8 time integrations, were vector-averaged along the time axis, leaving 1-dimensional datasets of 32768 channels. The data were then searched for visibilities with sufficient amplitudes to have an SNR $>$~5. Detected visibilities were then checked to see if they occur on all baselines. There were 200 signals satisfying these requirements, selected as narrow-band SETI candidates.

\subsubsection*{Frequency Distribution of the Candidate Signals}
\label{sec:freq_dist}

\begin{figure}[h]
\centering
\includegraphics[trim = 0mm 0mm 0mm 05mm, clip,scale=0.75]{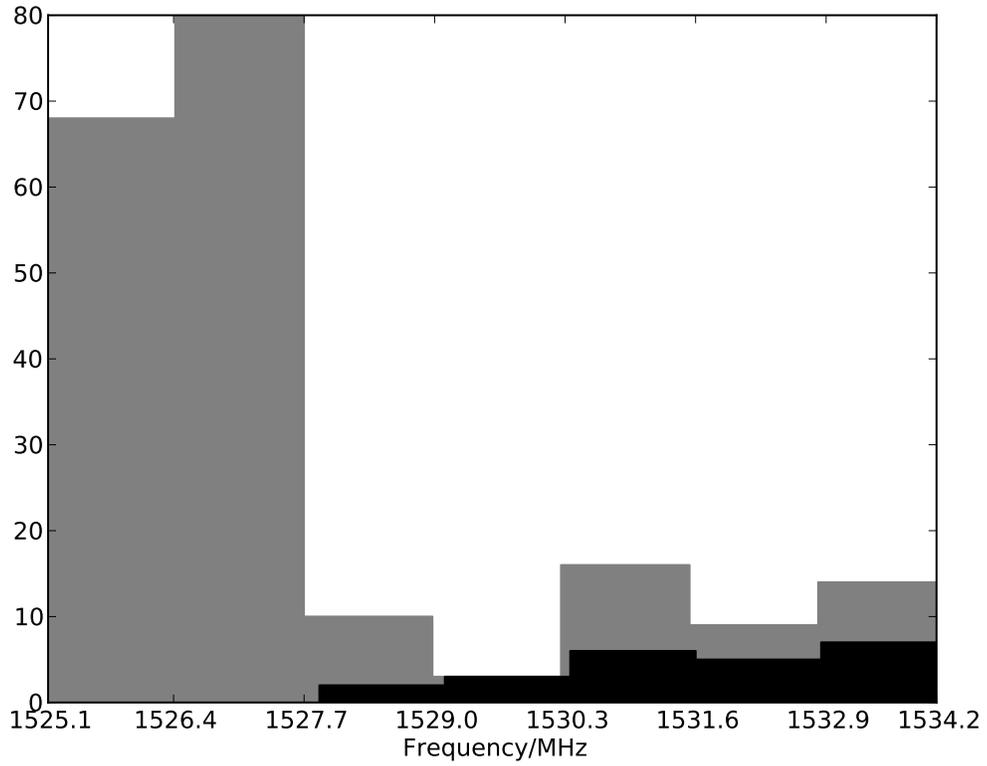}
\caption{Distribution of the SETI candidate signals, 200 narrow-band emission (grey) and 22 broad-band events (black). The SETI candidate signals fall within the operating frequency range (1525-1559~$\mega\hertz$) of the geostationary Australian Space-Earth transmission satellites, Inmarsat-Optus.}
\label{fig:allsig}
\end{figure}

\Fig~\ref{fig:allsig} shows the frequency distribution of the SETI candidate signals (1525-1534~$\mega\hertz$), with the 200 narrow-band (grey) and the 22 broad-band (black) candidate signals. Australian Space to Earth geostationary satellites, Optus' Mobilesat and the INMARSAT \footnote{The International Mobile Satellite Organisation (INMARSAT) is an international treaty organisation with sixty seven member countries including Australia. INMARSAT operates a GSO satellite system for global, mobile communications on land, sea and in the air} based systems, marketed in Australia by Telstra are known to operate at frequencies in the range  1525-1559~$\mega\hertz$ \citep{SDNTS}. The close match between the transmission frequencies of the satellites and our candidate signals suggests that many or all of our detected signals may be RFI coming from these satellites. 


\subsubsection{Phase Variation of broad-band Candidate SETI Signals}

\label{sec:freq_search}

Recall from \sect~\ref{sec:PhaseInfo}, $d\phi/d\nu$ is dependant on the location of the signal with respect to the phase centre. To obtain $d\phi/d\nu$ for the 22 broad-band SETI candidates, DFTs were applied to the complex visibilities as a function of frequency, as described in \sect~\ref{sec:FT}. The maximum expected $d\phi/d\nu$ for a radio signal from the planets is $(1.2\,\times\,10^{-10})$\deg per spectral channel (\sect~\ref{sec:phase_estimates}). This phase variation is effectively zero at the level of frequency sampling in the  observations.

\begin{figure}[h]
\centering
\includegraphics[scale=0.75]{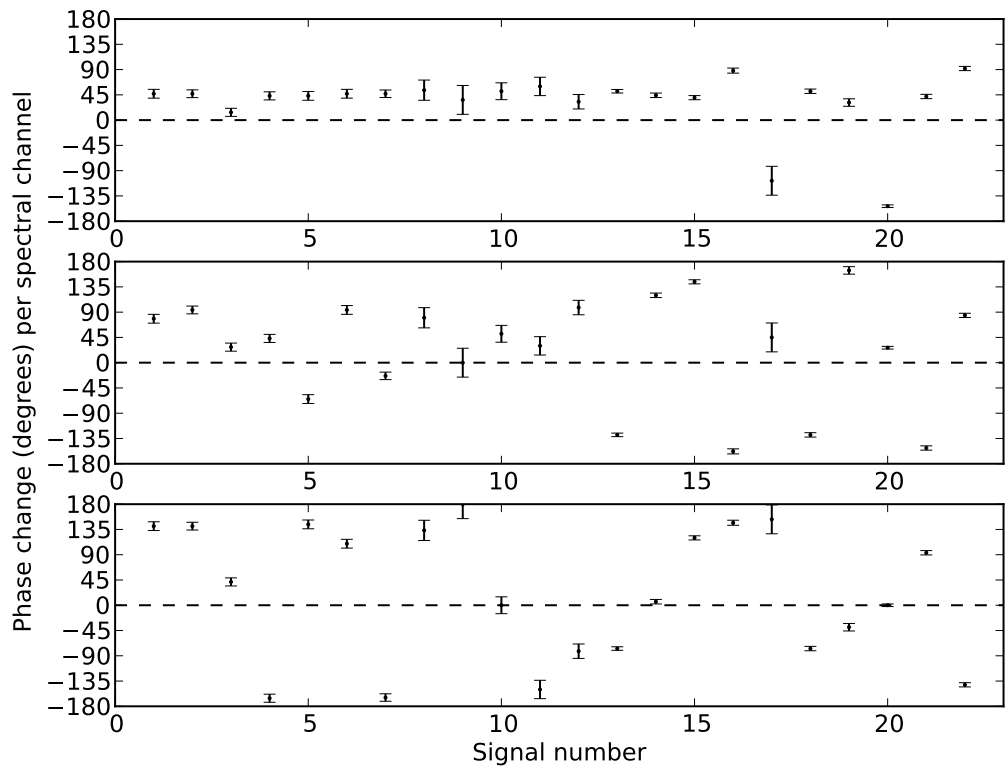}
\caption{The $d\phi/d\nu$ range and bin number for the 22 SETI broad-band candidate signals, for the three baselines is plotted on the y-axis (\textit{upper panel:} At-Mp; \textit{centre panel}: At-Pa; \textit{lower panel:} Mp-Pa). The error bars give the range of possible $d\phi/d\nu$ (or bin widths) for each signal. The x-axis gives the event number (1-22). The dashed line marks the maximum $d\phi/d\nu$ expected for a radio signal from Gl581, effectively zero at this resolution.}
\label{fig:phasefreq}
\end{figure}

\Fig~\ref{fig:phasefreq} shows the $d\phi/d\nu$ for the 22 broad-band SETI candidates. The horizontal dashed lines indicate the  $d\phi/d\nu$ range expected for a radio signal from the Gl581 system. Any signal with a $d\phi/d\nu$ range that crosses the horizontal dashed lines on all three baselines are considered as possibly originating from the Gl581 system, and merits further investigation. None of the 22 broad-band signals has satisfied this criteria, and are thus rejected as originating from the Gl581 planets. 

A large percentage of the signals are clustered around $d\phi/d\nu\,=\,45$\deg per spectral channel on the shortest baseline (At-Mp) in \fig~\ref{fig:phasefreq}. Using equation~\ref{eq:deltaphasefreq}, and assuming a North-South baseline of projected length, v = 0.5~$\mega\lambda$ (see \fig~\ref{fig:uvplot}),  this phase variation implies a minimum distance of $\sim$ 10\deg from the phase centre (c.f. declination of Gl581 $\simeq$ -7\deg, geostationary satellite declination $\simeq$ +5\deg). This effect is less clear on the longer baselines since the rapidly rotating phase of the complex visibilities along both the frequency and
time axes leads to averaging losses (smearing).


\subsubsection{Phase Variation of Narrow-Band Candidate SETI Signals}
\label{sec:time_search}

The discrete power spectra of the 200 narrow-band SETI candidate signals were obtained by application of DFTs to the complex visibilities as a function of time. 
As discussed in \sect~\ref{sec:phase_estimates}, the maximum $d\phi/dt$ expected from a radio signal originating from either of the planets is $2\,\times\,10^{-4}$~degrees per second. Such a slow varying signal is expected to behave as a signal from the phase centre, where most of its power appears in bin zero of the DFT power spectrum on the three baselines. 

To determine whether any of the signals are from the Gl581 system, classical detection theory was employed to test the hypothesis that the signal is located at the phase centre. Hypothesis testing compares the likelihood that any given signal is consistent with
a source located at the phase centre, to the likelihood that it is consistent with a source located away from the phase
centre, or from RFI. This is the classical \textit{Neyman-Pearson} (NP) approach to hypothesis testing, or signal detection \citep[chap.~3]{Kay1998}. The classical detection method has the advantages of providing robust estimation of errors and an efficient method of discriminating
false positives from true signals.


\subsubsection*{Hypothesis Testing and Monte-Carlo Simulations}

For Gaussian-distributed noise in the complex visibilities, the amplitudes of the DFT bins are Rician-distributed. The two hypotheses are as follows: (1) \textit{The alternative hypothesis}, $\mathcal{H}_{1}$; signal-present - a source located at the phase centre has most of the power concentrated in bin zero, yielding a Rice distribution of amplitudes with parameters (amplitude and variance) determined by the data SNR and noise level; (2) \textit{The null hypothesis}, $\mathcal{H}_{0}$; signal-absent - a source located away from the phase centre, yielding power in a bin other than bin zero, with the same Rician parameters. The ratio of the likelihoods for bin zero to the next highest bin is the test statistic.

\begin{figure}[h]
\centering
\includegraphics[scale=0.75]{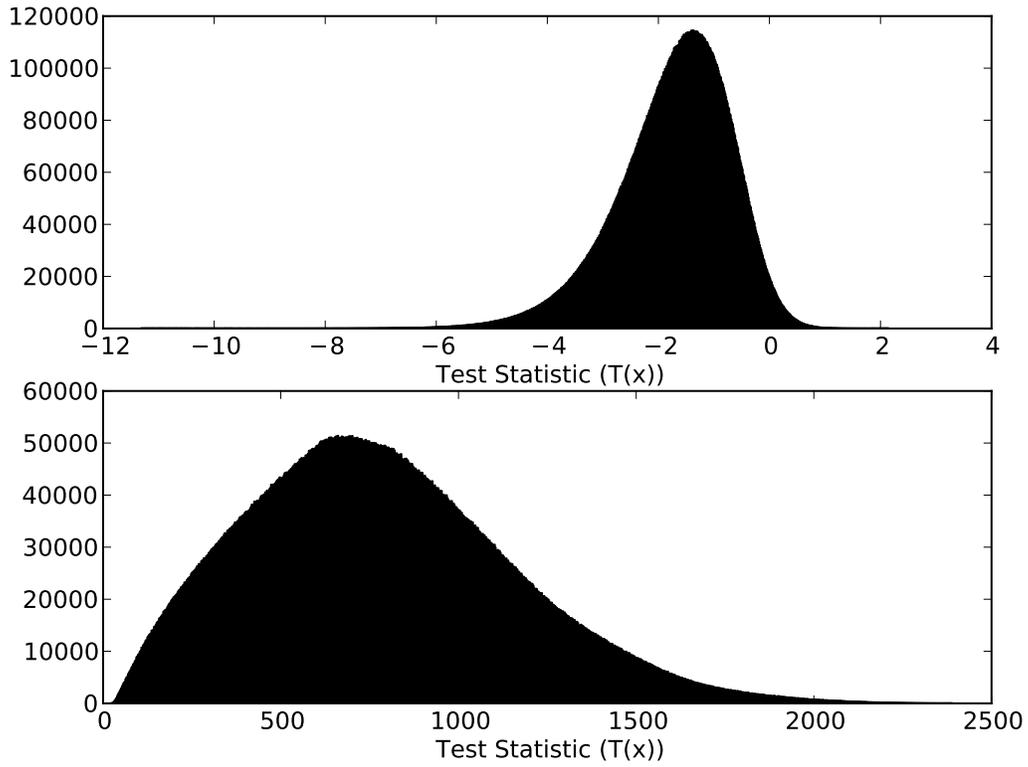}
\caption{Log-scale distribution of test statistics, $\ln[T(x)]$ obtained from Monte-Carlo simulations for both 1. Signal-absent (upper plot), and 2. Signal-present (lower plot). Note the very different ranges plotted on the x-axis.}
\label{fig:dpdt_hist}
\end{figure}

The test statistic is compared to a threshold to determine our level of confidence that the signal is real. To determine the threshold, we simulate $10^{6}$ signal-present and $10^{6}$ signal-absent sources, with the non-phase centre signals distributed randomly in the field, and for SNRs that reflect that found in our dataset. In addition, given that we expect the primary signals in the data to be RFI, rather than true signals, we also simulate zero-mean Gaussian noise with random RFI spikes located at a single time point in the visibilities. \Fig~\ref{fig:dpdt_hist} plots the log-scale distribution of the test statistics for both the signal-present (upper plot) and the signal-absent case (lower plot). The thresholds were chosen so as to minimise the chance of false-positives, i.e \textit{Type I error} \citep[chap.~3]{Kay1998}. From \Fig~\ref{fig:dpdt_hist} the thresholds ranges from the maximum log of the test statistic for the signal-absent case (= 2), and the minimum log of the test statistic for the signal-present case (= 35). 

\begin{figure}[h]
\centering
\includegraphics[scale=0.75]{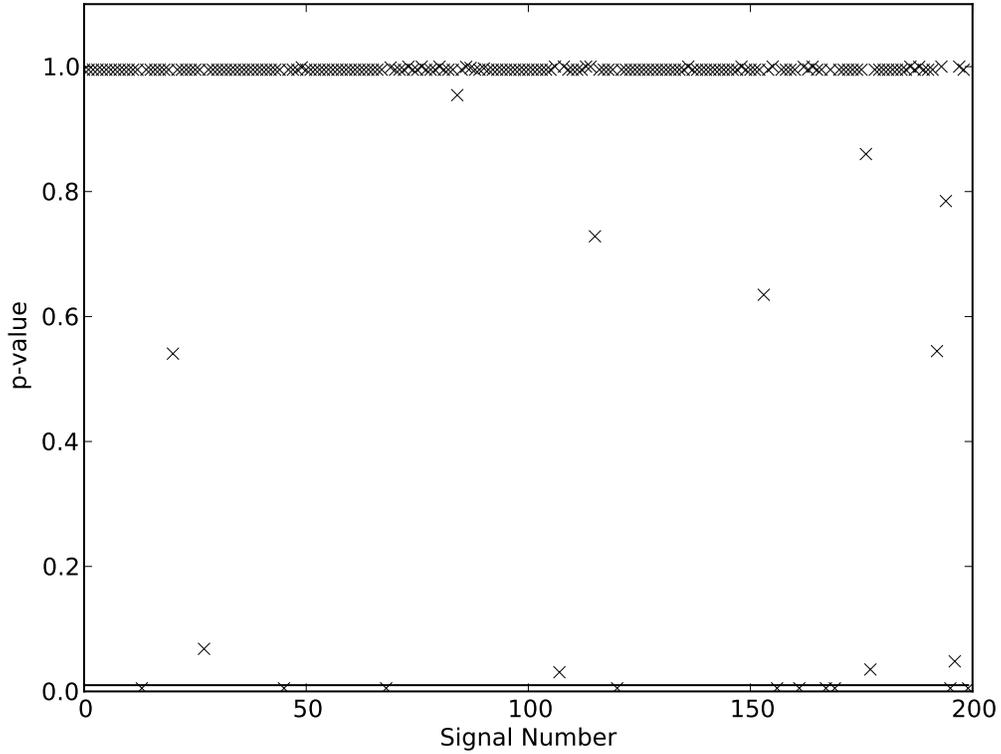}
\caption{Distribution of p-values for the 200 narrow-band signals. The p-values represent
the error in rejecting the null hypothesis, , $\mathcal{H}_{0}$ (signal-absent). The line plots the p = 0.01 significance value.
Signals at or below this line can be considered inconsistent with the signal-absent hypothesis, and require further analysis. The signals at p=1 all
have test statistics lower than the minimum test statistic of the simulated signal-absent distribution
(\fig~\ref{fig:dpdt_hist} ), and are assigned a p-value of 1.}
\label{fig:Pvalues}
\end{figure}

The method was then applied to the 200 narrow-band signals, where the overall test statistic, $T(x)$ was taken as the product of the test statistics of the individual baselines. Ten narrow-band signals were found to have $\ln[T(x)]\,>\,2$, of which only two are beyond  $\ln[T(x)]$ = 35 (\Tab~2). The data are expected to be consistent with no signal present at the phase centre (i.e. consistent with the null hypothesis). The significance level of rejecting the null hypothesis (the p-value) for the 200 signals were found by integrating the probability distribution of the signal-absent test statistic, obtained from the simulations. A small p-value ($p\,\leq\,0.01$) corresponds to a signal that may not be consistent with the null hypothesis, and warrants further investigation. \Fig~\ref{fig:Pvalues} shows the p-values of the 200 signals. Most of the 200 signals were found to have a p-value greater than 0.01, with ten signals having a p-value of $<$ 0.01.  

Eight signals lie between both distributions ($2\,<\,\ln[T(x)]\,<$~35). These signals are strictly inconsistent with both hypotheses, and reflect the limitations of our simple analysis; namely the simulation of data with either pure signals at the phase centre, or a single RFI peak in the visibility, and the need for the overall test statistic to have a value in excess of the threshold, rather than those for the individual baselines. 

\Tab~2 lists $\ln[T(x)]$ along with the test statistic for the individual baselines for all 10 events with $\ln[T(x)]\,>\,2$. For all but one of the 10 events, $\ln[T(x)]\,<\,2$ if the shortest baseline is excluded. For all 10 events the test statistic is far greater for the shorter baseline than for the other baselines. This is not consistent with a source at the phase centre. It is, however, consistent with RFI distant from the phase centre. We can therefore easily exclude these signals as being from the GL581 system.


%


\section{Discussion}
\label{sec:Discussion}
We present the first VLBI SETI experiment. This pilot study demonstrates that the fundamental properties of VLBI makes it an ideal technique for targeted SETI. With multiple baselines, milliarcsecond resolution, and millijansky sensitivities, RFI can be easily identified and rejected using automated techniques, while being sensitive to very weak signals.

Our search method targeted signals that are five times the baseline sensitivity, which were then cross-referenced over three baselines, to identify corresponding signals. Hence, the amplitude sensitivity of our array was limited to our least sensitive baseline over an 8$\times$3 minute observing interval, for narrow-band signals (At-Mp; $1\sigma = 0.31$~mJy) and a single 3 minute scan for broad-band signals (a factor of $\sqrt{8}$ less sensitive)\footnote{The sensitivity limits were derived using the System Equivalent Flux Densities (SEFD) for the individual telescopes at 1400~$\mega\hertz$; At=68~Jy, Pa=40~Jy, Mp=340~Jy.}.

Our experiment finds no evidence for radio signals $>\,1.55~\milli$Jy (= $5\sigma$) within the frequency range of 1230 - 1544~$\mega\hertz$ from the region of Gl581. We can therefore place an upper limit of $7~\mega\watt\,\reciprocal\hertz$ on the power output of any isotropic emitter located in this planetary system, within this frequency range. On the other hand the power of the Arecibo planetary radar transmitter is $\sim\,1\,\mega\watt$ \citep{Black2002, Tarter04}. Assuming an aperture efficiency of 50$\%$, at $\lambda\,=\,21\,\centi\metre$ the directive power output is $\sim$6$~\tera\watt$. Over a bandwidth of 5.4~$\hertz$ \citep{Blacketal2001}, this emission would have the flux density of a $\sim$200~Jy radio source (or 650~mJy if integrated over our channel width of 1.93~$\kilo\hertz$) at a distance of 20~ly.

\subsection*{Using current VLBI instruments for SETI}

Since this was a proof-of-concept observation, two differences between this experiment and a typical VLBI observation should be noted.
Firstly the baseline lengths in this experiment were short by VLBI standards. 
Baselines one or even two orders of magnitude longer would significantly reduce the amount of correlated RFI.
Secondly, most VLBI observations are carried out using more than three telescopes.
Increasing the number of baselines to four or more constrains both the phase and amplitude closure relations\footnote{The sum of the phases on any triangle of three baselines, and the combined amplitude ratio of four baselines contains information only on the true
visibility of the source itself, all other errors cancel out \citep[chap.~9]{TMS}.}, drastically increasing the quality of calibration. This would also provide the opportunity for image plane searching. However, it would likely be computationally expensive as a technique for searching over very wide frequency ranges.
Relaxing the single baseline five times median limit could enhance the sensitivity beyond that of the weakest baseline.

We should also note that the width of the DFT bin is not a fundamental limit on the determination of the phase variation.
A least-squares fit would allow the delay and delay rate to be determined more accurately.
Essentially, the error on these parameters is limited by the signal to noise.
Certainly the LBA has been used for astrometry at the sub-milliarcsecond level \citep{Dellaretal09a,Dellaretal09b}.
Observations of spacecraft have proven VLBI astrometry techniques for very weak narrow-band sources \citep{APG06} and suggest that VLBI might even allow the detection of the orbital motion of SETI signals.

While the techniques discussed in this paper are ideal for targeted SETI, we also note the possibility of carrying out SETI commensally with normal VLBI operations.
SETI target stars \citep{TT03} would be expected to be present in the field of view of the antennas for VLBI observations reasonably regularly.
Using the techniques discussed in \citet{Delleretal11} and \citet{Morganetal11} it would be possible to produce visibility datasets for any target star in the primary beam. Additionally, direct analysis of the voltages (the baseband data) although computationally expensive, could also be implemented at the correlator in an efficient manner to search for signals from extraterrestrial intelligence.
A possible model for this is the VFASTR transient search currently running at the VLBA operations centre \citep{Waythetal11,Thompsonetal11}.

\subsection*{SETI and the Square Kilometre Array (SKA)}

A sensitive and advanced radio telescope, the Square Kilometre Array (SKA) is
currently being planned for the coming decades\footnote{\url{http://www.skatelescope.org/about/project/}}.

At 1400~$\mega\hertz$, the long baselines of the SKA will provide an astrometric precision of $\sim$3~$\mu$-arcseconds \citep{FR04}.
In addition, the SKA is expected to achieve sub-$\mu$Jy sensitivities \citep{CR04}.
At $\sim$20~ly the SKA would probe an isotropic luminosity of a few $\kilo\watt\reciprocal\hertz$, far below the power output of radars and telecommunication satellites \citep{Tarter04}.
The fields of view will be $\sim$1 square degree at the maximum baselines and $\sim$30 square degrees for the shorter baselines \citep{CR04}.
These capabilities will make the SKA the fastest, most sensitive and widest field of view survey instrument in astronomy.
However, for the SKA to be used for SETI, suitable targets are required.
Such targets are currently being provided by the Kepler Mission \citep{Kepler2, Kepler3}.

\section{Conclusion}
The discoveries of potential habitable planets are no doubt fuelling a renewed excitement in SETI.
With current VLBI systems we are able to search these candidates for signals from extra-terrestrial civilizations over a wide range of frequencies, and with high sensitivity.

In addition, VLBI is undergoing a transition from traditional recording systems towards real-time correlation via network connections (so called e-VLBI). 
This has led to VLBI being used for rapid follow-up of transient sources.
These developments, the wide range of frequencies covered by VLBI instruments, and the huge added value that VLBI techniques bring to a targeted SETI search make VLBI the natural choice for follow-up observations to confirm a potential SETI signal. However, more accurate positions are needed for target stars for future experiments with higher angular resolution. Such positions will be provided in the near future by the high-quality optical astrometry of \textit{Gaia} \citep{GAIA}.

\acknowledgments

The International Centre for Radio Astronomy Research is a joint venture between Curtin University and the University of Western Australia, funded by the state government of Western Australia and the joint venture partners. The Long Baseline Array is part of the Australia Telescope which is funded by the Commonwealth of Australia for operation as a National Facility managed by CSIRO. SJT is a Western Australian Premier's Research Fellow, funded by the state government of Western Australia. The Centre for All-sky Astrophysics is an Australian Research Council Centre of Excellence, funded by grant CE11E0090. We wish to thank Adam T. Deller for performing the correlation of the LBA dataset, and James Miller-Jones for assisting in the calculation of the positional offset of Gliese~581.

{\it Facilities:} \facility{Long Baseline Array (ATCA, Mopra, Parkes)}





\clearpage

\begin{table}[h]
\centering
\begin{tabular}{|c|c|}
\hline
UT& Frequency Pairs ($\mega\hertz$)\\
\hline 
07:40:00-08:30:00 & 1262,1312\\ \hline
08:30:00-09:10:00 & 1362,1412 \\ \hline 
09:10:00-09:50:00 & 1462,1512 \\ \hline
09:50:00-10:30:00 & 1262,1312 \\ \hline
10:30:00-11:10:00 & 1362,1412 \\ \hline
11:10:00-11:50:00 & 1462,1512 \\ \hline
11:50:00-12:30:00 & 1262,1312 \\ \hline
12:30:00-13:10:00 & 1362,1412 \\ \hline
13:10:00-13:50:00 & 1462,1512 \\ \hline
13:50:00-14:30:00 & 1262,1312 \\ \hline
14:30:00-15:10:00 & 1330,1380 \\ \hline
15:10:00-15:30:00 & 1462,1512 \\ \hline
\hline
\end{tabular} 
\medskip
\caption{The observation was sub-divided into 12 separate sessions, that alternated between 3 frequency pairs. This table lists the observation times and the corresponding frequency pairs.}
\label{tab:table1}
\end{table}

\begin{table}[h]
\begin{tabular}{|c|c|c|c|c|c|}
\hline
\multicolumn{1}{|c}{Frequency} & \multicolumn{1}{|c}{UT} & \multicolumn{1}{|c}{$\ln[T(x)]$} & \multicolumn{3}{|c|}{Baseline Specific log of the test statistic} \\ \hline
  \multicolumn{1}{|c}{(MHz)} & \multicolumn{1}{|c}{} & \multicolumn{1}{|c}{} & \multicolumn{1}{|c}{At-Mp} & \multicolumn{1}{|c}{At-Pa} & \multicolumn{1}{|c|}{Pa-Mp} \\ \hline
\multicolumn{1}{|c}{1531.69080115} & \multicolumn{1}{|c}{13:10:00-13:50:00} & \multicolumn{1}{|c|}{2.39} & \multicolumn{1}{|c}{1.92} & \multicolumn{1}{|c}{-0.30} & \multicolumn{1}{|c|}{0.77} \\ \hline
\multicolumn{1}{|c}{1533.6297203} & \multicolumn{1}{|c}{09:10:00-09:50:00} & \multicolumn{1}{|c}{3.11} & \multicolumn{1}{|c}{6.12} & \multicolumn{1}{|c}{-1.36} & \multicolumn{1}{|c|}{-1.65} \\ \hline
\multicolumn{1}{|c}{1533.80206867} & \multicolumn{1}{|c}{13:10:00-13:50:00} & \multicolumn{1}{|c}{3.50} & \multicolumn{1}{|c}{2.19} & \multicolumn{1}{|c}{0.66} & \multicolumn{1}{|c|}{0.66} \\ \hline
\multicolumn{1}{|c}{1527.85017443} & \multicolumn{1}{|c}{09:10:00-09:50:00} & \multicolumn{1}{|c}{4.21}& \multicolumn{1}{|c}{4.89} & \multicolumn{1}{|c}{1.40} & \multicolumn{1}{|c|}{-2.07} \\ \hline
\multicolumn{1}{|c}{1527.50351919} & \multicolumn{1}{|c}{11:10:00-11:50:00} & \multicolumn{1}{|c}{6.51} & \multicolumn{1}{|c}{16.91} & \multicolumn{1}{|c}{-4.95} & \multicolumn{1}{|c|}{-5.45} \\ \hline
\multicolumn{1}{|c}{1527.56227431} & \multicolumn{1}{|c}{11:10:00-11:50:00} & \multicolumn{1}{|c}{7.25} & \multicolumn{1}{|c}{11.04} & \multicolumn{1}{|c}{-0.27} & \multicolumn{1}{|c|}{-3.52} \\ \hline
\multicolumn{1}{|c}{1525.18856723} & \multicolumn{1}{|c}{09:10:00-09:50:00} & \multicolumn{1}{|c}{14.84} & \multicolumn{1}{|c}{30.53} & \multicolumn{1}{|c}{-15.32} & \multicolumn{1}{|c|}{-0.37} \\ \hline
\multicolumn{1}{|c}{1527.50156068} & \multicolumn{1}{|c}{09:10:00-09:50:00} & \multicolumn{1}{|c}{26.19} & \multicolumn{1}{|c}{17.86} & \multicolumn{1}{|c}{2.87} & \multicolumn{1}{|c|}{5.47} \\ \hline\hline
\multicolumn{1}{|c}{1532.03158088} & \multicolumn{1}{|c}{09:10:00-09:50:00} & \multicolumn{1}{|c}{86.19} & \multicolumn{1}{|c}{88.270} & \multicolumn{1}{|c}{0.03} & \multicolumn{1}{|c|}{-2.10} \\ \hline
\multicolumn{1}{|c}{1527.56423282} & \multicolumn{1}{|c}{09:10:00-09:50:00} & \multicolumn{1}{|c}{242.77} & \multicolumn{1}{|c}{1782.83} & \multicolumn{1}{|c}{-1305.76} & \multicolumn{1}{|c|}{-234.30} \\
\hline
\end{tabular}
\medskip
\caption{The ten signals with $\ln[T(x)]\,>\,2$.
\textit{Column 1}: The centre frequency; \textit{Column 2}: The observed (UT) time period; \textit{Column
3}: The sum of the log of the test statistics of the individual baselines, which is taken as the final test
statistic; \textit{Columns 4-6}: The log of the test statistics of the individual baselines.}

\label{tab:table2}
\end{table}


\end{document}